\documentclass[%
 reprint,
 amsmath,amssymb,
 aps,
]{revtex4-2}

\usepackage{graphicx}
\usepackage{dcolumn}
\usepackage{bm}
\usepackage{float}
\usepackage{xcolor}
\usepackage{soul} 
\usepackage{siunitx}

\begin{document}

\title{Interplay of Mechanochemistry and Material Processes in the Graphite to Diamond Phase Transformation}

\author{Brenden W. Hamilton}
 \email{brenden@lanl.gov}
\author{Timothy C. Germann}
 \affiliation{
 Theoretical Division, Los Alamos National Laboratory, Los Alamos, New Mexico 87545, USA}

\date{\today}

\begin{abstract}

The manifestation of intra-molecular strains in covalent systems is widely known to accelerate chemical reactions and open alternative reaction paths. This process is moderately understood for 
isolated molecules and uni-molecular processes.
However, in condensed matter processes such as phase transformations, material properties and structure may influence typical mechanochemical effects.
Therefore, we utilize steered molecular dynamics to induce out of plane strains in graphite and compress the system under a constant strain rate to induce phase transformation.
We show that the out of plane strain allows for phase transformations to initiate at lower amounts of compressive strain.
Yet, in contrast to typical mechanochemical results, the sum of compressive and out of plane work needed to form diamond has a local minimum due to altered defect formation processes during phase transformation.
Additionally, these altered processes slow the kinetics of the phase transformation, taking longer from initiation to total material transformation.

\end{abstract}

\maketitle

Mechanochemistry is known to accelerate chemical reactions and alter the reaction path in covalent systems \cite{Ribas-Arino2012Review,Takacs2013Historical,Stauch2016Review,Hamilton2022Extemporaneous}. 
By adding additional strain to chemical bonds, the activation barrier can be directly lowered \cite{Bell1978Models}, typically resulting in faster reaction kinetics.
Mechanochemistry can occur in cases in which a mechanical strain drives a bond to yield and induces chemistry \cite{Grandbois1999HowStrong,Davis2009ForceInduced}, as well
as mechanochemical assistance on a thermally induced process of interest \cite{Hamilton2021HotspotsBetterHalf,Wood2015UltrafastChemistry,Hamilton2022Extemporaneous,Hamilton2022PEHotspot,Hamilton2021Review,hamilton2023energy}.

While the majority of mechanochemistry research has focused on the linear strains induced by pulling on covalent bonds \cite{Ribas-Arino2009Understanding,Davis2009ForceInduced,Ong2009FirstPrinciples,Ghanem2021Role,Wang2015Inducing,Deneke2020Engineers,Stauch2014JEDI,Wiita2006ForceDependent,Dopieralski2013JamusFaced,Piermattei2009Activating}, 
it has recently been shown that the
dynamics of condensed phase, molecular materials under complex strain states can induce 'many-body' deformations: the twisting and bending of molecules \cite{Hamilton2021HotspotsBetterHalf,hamilton2022rapid}. These many-body deformations can induce
significant acceleration of reactions via lowered energy barriers and cause alternate reaction pathways to become the main reaction path \cite{Hamilton2022ManyBody,Kroonblawd2020ShearBands,Larsen2013FlexActivated,Klukovich2013Backbone}. 
In the pulling of mechanophores to induce isomerization, adding a rotational strain around the central axis of the molecule was shown to lower the total energy needed to cause reaction. This monotonic decrease in the \emph{total work} needed, 
the pulling work done and the energy needed to induce the rotational strain, results in a more efficient reaction path than pure pulling forces \cite{Hamilton2022ManyBody}.

Layered materials, such as graphite, are known to have layer buckling instabilities under pressure, which can induce local, out of plane strains \cite{Lafourcade2018Irreversible,lafourcade2020elastic,plummer2022basal,barsoum2011elastic}. Specifically in graphite, these local
deformations can close the HOMO-LUMO gap to potentially assist in reactivity and phase transformation \cite{Kroonblawd2018MechanochemicalGraphite}. 
Graphitic defects such as ripplocations\cite{badr2022ripplocations,barsoum2017deformation,barsoum2019ripplocations,freiberg2018nucleation}, kinking\cite{barsoum2004kink,plummer2021origin,basu2009spherical}, 
and Wigner defects\cite{telling2003wigner,ewels2003metastable} can all result in a wide range of out of plane strain states of a layer.
Carbon phase transformations and diamond formation have significance for a variety of scientific topics such as
detonation products \cite{hammons2021submicrosecond,Armstrong2020Ultrafast}, planetary interiors \cite{hubbard1981interiors,kraus2017formation}, nuclear fusion \cite{biener2009diamond}, and synthetic materials fabrication \cite{balmer2009chemical}. Rapid
compression and shear stress formation have been commonly proposed as a viable route for the synthetic formation of diamond \cite{Kraus2016Nanosecond,scandolo1995pressure,hirai1991modified}.
The open questions are: (a) how does local out of plane strain influence the graphite
to diamond phase transformation, beyond typical pressure effects; and (b) can it be utilized to tailor nanoscale diamond formation,  better controlling other carbon allotropes? More broadly, previous many-body mechanochemistry studies have looked at isolated molecules and uni-molecular decomposition reactions. In the case of graphite to diamond phase transformations, 
significantly more condensed matter processes are in play, such as defect formation and non-hydrostatic stress states which will more directly influence the transformation kinetics.

In this work, we utilized Many-Boded Steered Molecular Dynamics (MBsMD) \cite{Hamilton2022ManyBody} to induce out of plane strain on graphite layers, followed by a uniaxial compression to induce phase transformation.
All molecular dynamics (MD) simulations were conducted using the LAMMPS software package \cite{Plimpton1995LAMMPS,thompson2022lammps}. Interatomic forces were calculated with the ReaxFF potential \cite{Wood2018Multiscale}. All simulations were conducted under 
NPT conditions with fully periodic boundary conditions and a 0.1 fs timestep.
The temperature was set to 300 K using a Nose-Hoover thermostat \cite{Nose1984Unified} and pressure conditions were set to 1.0 atm.
The X and Y pressure tensor components are coupled, and all shear stresses (box tilts) are independent of each other. Uniaxial deformations are applied in the 
Cartesian Z direction with a constant engineering strain rate. Deformation is performed at every timestep and atomic coordinates are fractionally remapped.

\begin{figure*}[ht]
  \includegraphics[width=0.9\textwidth]{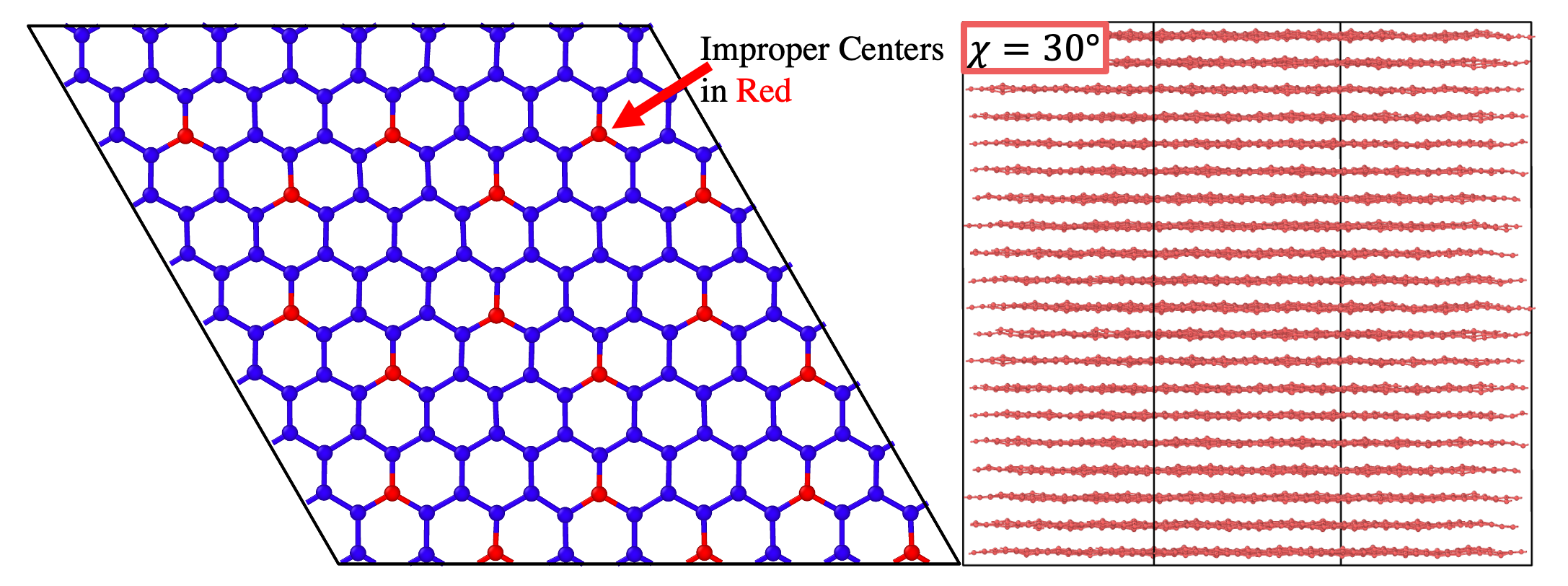}
  \caption{Initial simulation set up. Left: top-down view, Z axis into page, of graphite layers with improper centers colored red. Improper centers in the adjacent layer are shifted by one carbon ring. Right: side-on view of an out of plane strained system with 
  the set angle as 30$^{\circ}$.}
  \label{fig:Fg1}
\end{figure*}

The initial atomic configuration is a hexagonal graphite cell of 12960 atoms and (300 K, 1 atm) relaxed box dimensions of X = 4.56918 nm, Y = 3.95348 nm, and Z = 6.58632 nm, with the non-zero box tilt of XY = -2.28255 nm. MBsMD is used to apply an external field to drive 4-body systems
(either proper or improper dihedrals) to be at specific angles, causing twisting and bending deformations within molecules \cite{Hamilton2022ManyBody}. Here, improper dihedrals within the graphite layers are used to induce out of plane strain. 
Figure 1 colors the improper dihedral center atoms red for one graphite layer. Improper angles, which are 0$^{\circ}$ in the relaxed crystal, are driven to angles between
0$^{\circ}$ and 50$^{\circ}$, with independent runs for a chosen angle.
Figure 1 shows a side on view of layer impropers driven to 30$^{\circ}$. Before compressive deformation,
the cells are equilibrated with the external field on for 50 ps at 300 K and 1 atm. For ensembled samples, an additional 20 ps was run, saving a configuration every 5 ps.

In this work, the MBsMD method is extended such that the external field is only applied to a specific improper if its bonding environment is in its initial state. Hence, the chosen angle does not directly deform or alter the diamond phase regions. 
Additional details are provided in Supplemental Material section SM-1.

The phase of the material is determined using the polyhedral template matching as employed in LAMMPS \cite{larsen2016robust}. As shown in Figure 2, the strains (or times) of initiation and completion of transformation from graphite are defined as the first step in which the graphite population
fails below 80\% and 20\%, respectively. The transformed phase populations are taken for all systems as the state at 50\% strain.
The compressive work is calculated from the internal energy rise from the first step of compression to the peak energy state prior to phase transformation. 
The total work is calculated using the same energy peak, but referenced from the initial state with no external field system, i.e., the combined work done from compression and to initially deform the layers.

\begin{figure}[h]
  \includegraphics[width=0.4\textwidth]{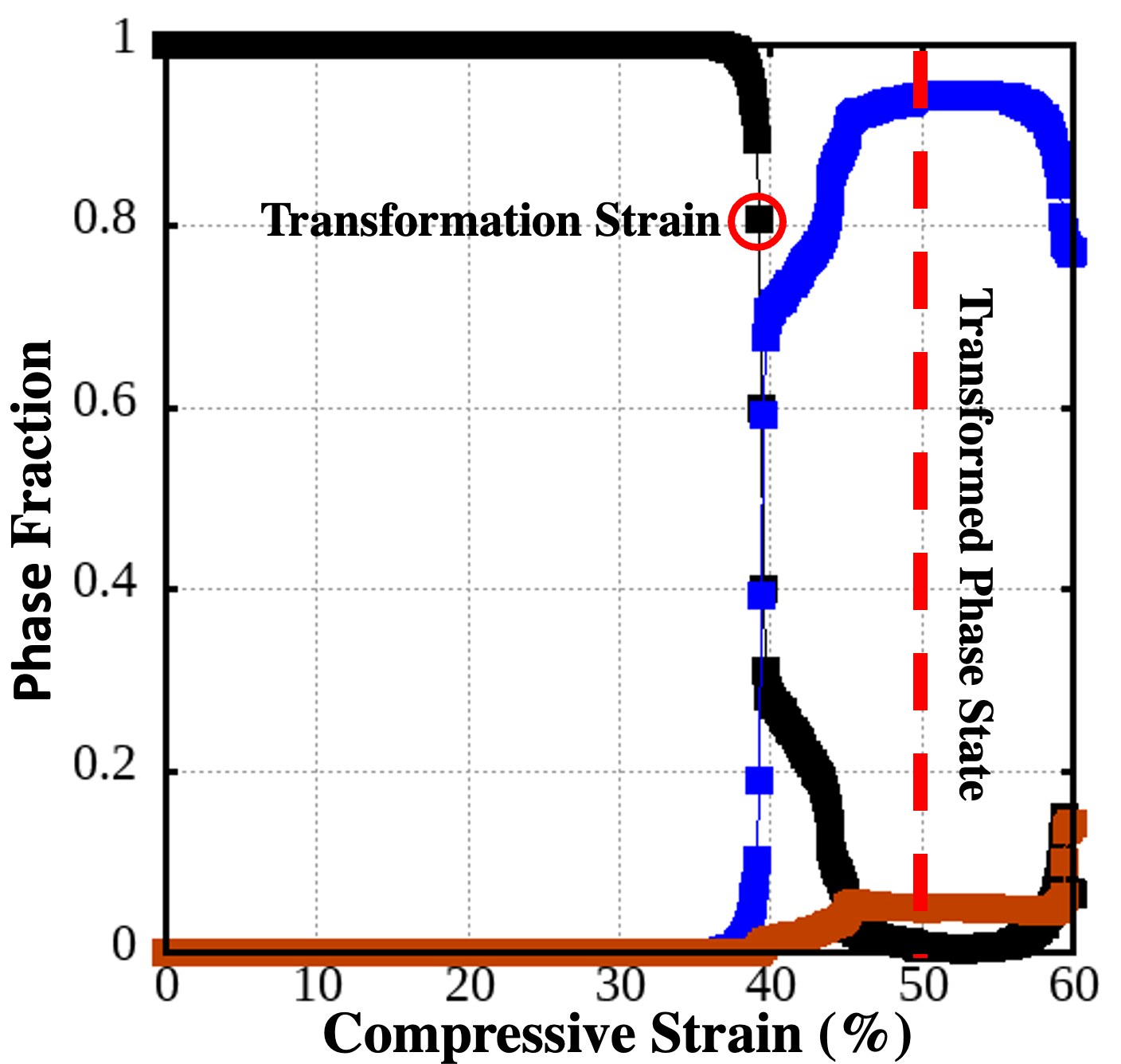}
  \caption{Example of phases present during compression for a case with out of plane strain at 30$^{\circ}$ and a strain rate of \num{5e-9}/s. Black points are graphite, blue for cubic diamond, orange for hexagonal diamond. Circle at
  40\% strain and 0.8 population represents the point of phase transformation used in Figure 3. The vertical line at 50\% strain indicates the point in which phase amounts are taken for Figure 4.}
  \label{fig:Fg2}
\end{figure}

For each initial out of plane strain, 5 samples are compressed with a constant engineering strain rate of \num{5e-9}/s. Figure 3 shows the strain at which transformation initiates (blue) and finishes (red). 
The inclusion of an out of plane strain directly lowers the compressive strain necessary to begin to form a diamond phase, with out of plane improper angles of 50$^{\circ}$ inducing the phase transformation at $\approx$10\% strain lower than at 0$^{\circ}$ out of plane strain. While the average points show
a monotonic decrease, there is still considerable noise from sample to sample, as expected from previous works \cite{Kroonblawd2018MechanochemicalGraphite}. Figure 2, which shows the phases present for 30$^{\circ}$ out of plane strain, shows a rapid transformation from
100\% graphite to $\approx$35\% graphite, followed by a much slower completion of the transformation to a fully diamond state (mixture of cubic and hexagonal). At higher out of plane strains,
small regions of a locally disordered lattice form as meta-stable regions that often become diamond after further compression, leading to more compression/time overall to complete the phase transformation to a (mostly) diamond system.
This results in the divergence of the 80\% graphite strain (blue) and 20\% graphite strain (red) curves, which represents a slowing of the phase transformation kinetics.
While the increased out of plane strain lowers the activation barrier for phase transformation, it also results in additional materials processes
such as defect formation that slows the overall kinetics, especially as the systems reach lower amounts of graphite remaining. We devote the remainder of this paper to understanding the processes underlying the slower reaction kinetics.

\begin{figure}[ht]
  \includegraphics[width=0.4\textwidth]{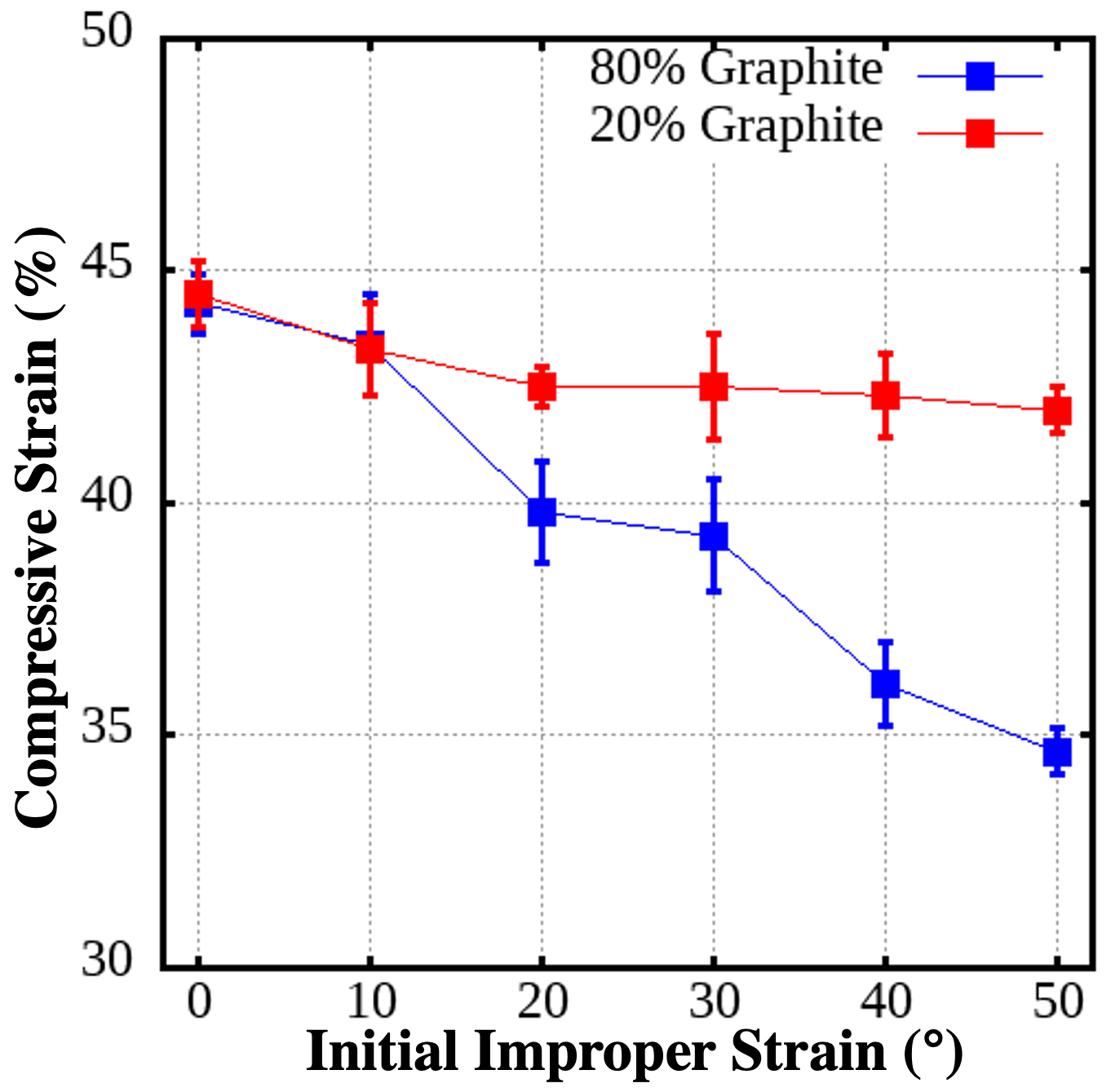}
  \caption{Level of compressive strain needed to induce (blue, 80\% graphite) and complete (red, 20\% graphite) phase transformation for out of plane strains studied. 
  Points are averaged over five independent runs with error bars as standard deviation.}
  \label{fig:Fg3}
\end{figure}

To better assess the overall process of diamond formation under an induced out of plane strain,
Figure 4 shows the average work done on the system to induce phase transformation for each out of plane strain, where
the total work includes both the compressive work as well as the work done to create the out of plane strain states. For no out of plane strain, these values are the same, and a monotonic decrease is expected for increasing strain,
as was shown for the activation barriers of decomposition reactions \cite{Hamilton2022ManyBody}. The increasing divergence of the two is also expected as more work is done to reach higher levels of out of plane strain.
The total work shows an uptick for out of plane strains from 20$^{\circ}$ to 40$^{\circ}$, which is accompanied by a greatly reduced slope for the compressive work and continual increasing out of plane strain work, resulting in a local minimum at 20$^{\circ}$. 
While the error bars show that this uptick is not necessarily statistically significant, this trend of a slowed decrease in the work necessary may physically arise from the onset of structural damage. This is consistent with the slower transformation kinetics for the last 30-35\% of graphite transformation observed for higher out of plane strains (Figs.~2 and 3), since additional compressive work is needed to complete the  transformation.
The 20$^{\circ}$ represents a local minimum where
significant local disorder has yet to significantly form, leading to minimal slowing of kinetics and no effect on energetics, letting the system benefit from a large mechanochemical effect from the out of plane strain, without it impeding the transformation. 
The Supplemental Material section SM-2 shows a graphite population history for each out of plane strain, as well as mid-transformation images of atomic positions, from snapshots as close to the onset of local disorder as possible (between 30\% and 50\% graphite population) when relevant. 

These results differ from previous studies of the isomerization of mechanophores, in which
adding a rotational strain prior to pulling on the molecule to induce reaction consistently lowered the total work needed for reaction \cite{Hamilton2022ManyBody}.
This helps to showcase the complex nature of mechanochemistry in condensed matter systems (compared to isolated molecules and 
idealized systems) in which deviatoric stresses, local phase and structure, defect formation, and the existence of multiple meta-stable states all play a role in how the reactivity of the system responds. In this case, too much mechanochemical influence
ends up altering other processes after the onset of phase transformation that partially negates the initial acceleration.

\begin{figure}[ht]
  \includegraphics[width=0.4\textwidth]{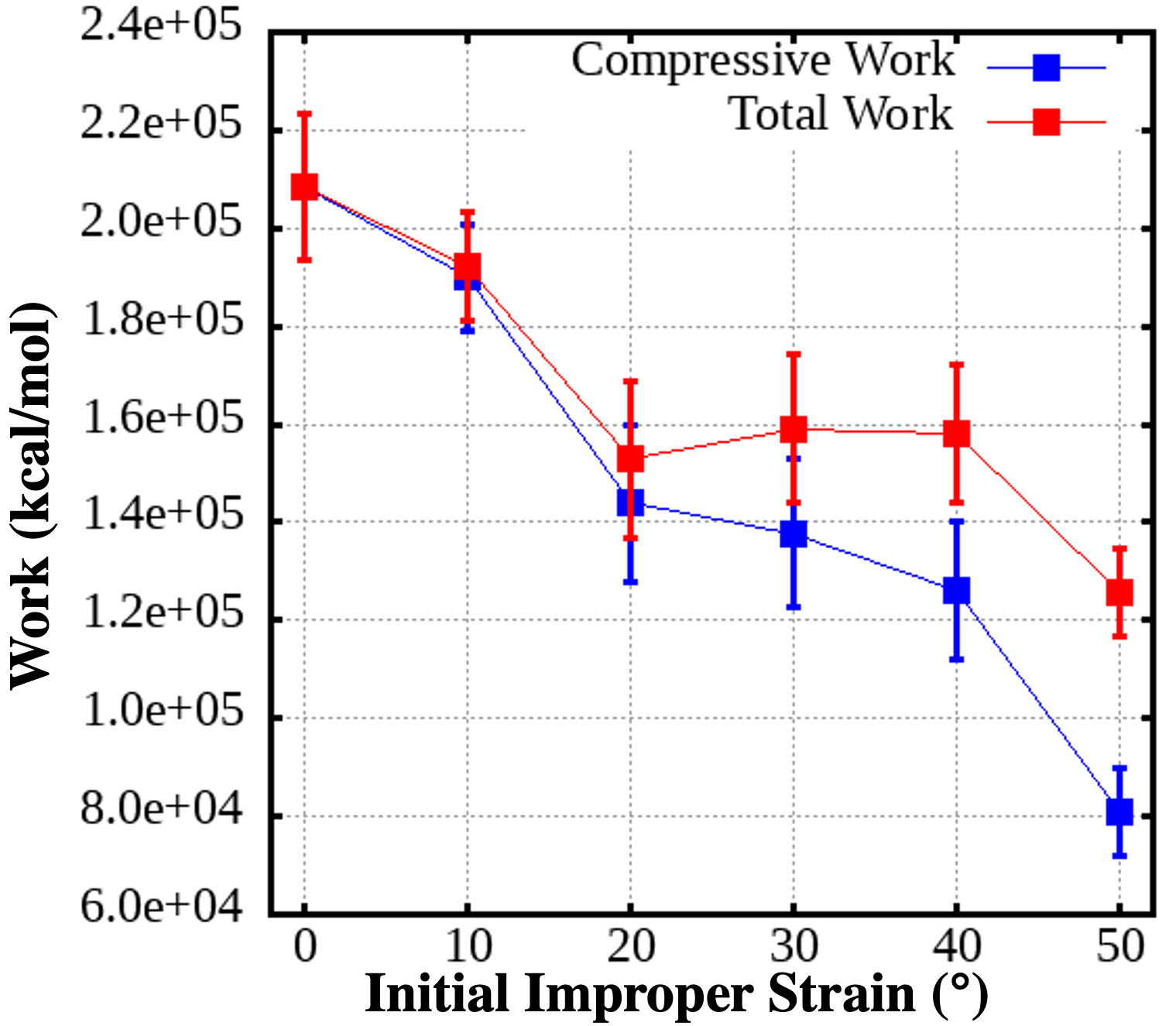}
  \caption{Work done on the system prior to phase transformation. Compressive work (blue) is from just uniaxial deformation. Total work (red) includes compressive work and out of plane strain work. 
  Points are averaged over five independent runs Error bars represent the standard deviation from five independent runs.}
  \label{fig:Fg5}
\end{figure}

Finally, Figure 5 shows the phase populations at 50\% strain, in which graphite has typically transformed to a majority of some phase of diamond, either cubic or hexagonal. The hexagonal diamond
formed in the low out of plane strain cases is from the formation of pure twin defects. With increasing out of plane strain, the amount of hexagonal diamond, as well as the uncertainty, in phase amounts, increases. This is due to both the presence of twin defects as well
as local disorder and dislocations which can manifest as hexagonal diamond under continued compression that results in crystallization.
Supplemental Material section SM-3 shows these same population plots for 45\% and 55\% compressive strain, in which the same trends arise where 20$^{\circ}$ out of plane strain is the local maximum of cubic diamond and uncertainty greatly increase
beyond this point. This helps to corroborate the idea that the local minimum in work done is a result of material influences on kinetics coupled with the mechanochemical effect and not an artifact of an arbitrary
choice in total strain to assess populations, which are moderately stable from transformation until a compression induced material failure.
The local minimum in total work at 20$^{\circ}$ may also arise from the lack of pure twin formation at this level of out of plane strain due to local energetics and/or available transition states.

\begin{figure}[htb]
  \includegraphics[width=0.4\textwidth]{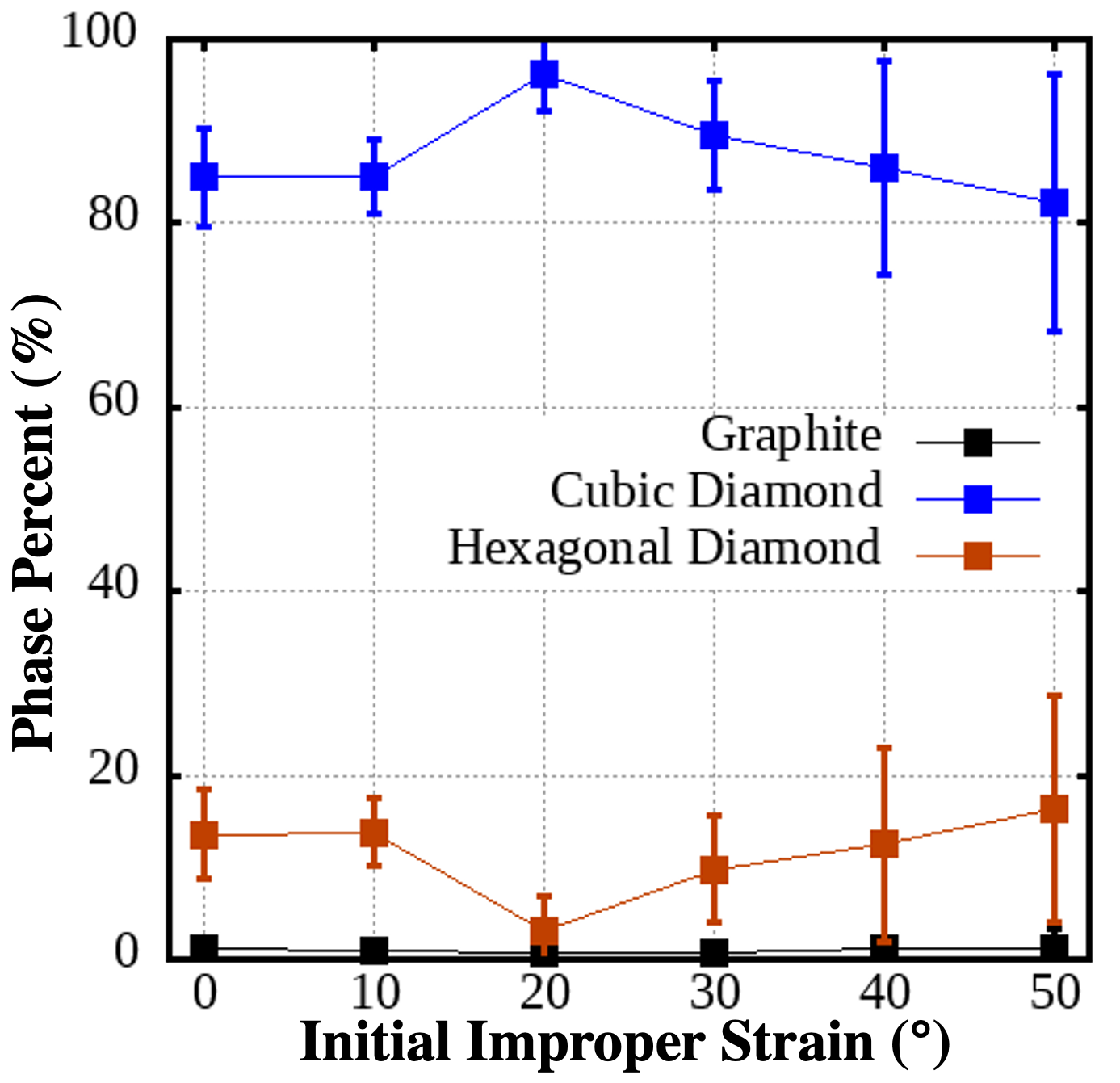}
  \caption{Percentage amounts of each phase present at 50\% compressive strain. Points are averaged over five independent runs, and error bars represent the standard deviation over those runs. 
  Black points are graphite, blue for cubic diamond, orange for hexagonal diamond.}
  \label{fig:Fg4}
\end{figure}

In summary, we assess how mechanochemical perturbations of the graphite structure influence the compression induced transition to diamond. By using steered MD to induce out of plane strains, systems with more out of plane strain require less total compression
to form diamond, in terms of total strain, but the overall kinetics of total transformation are considerably slower. Assessing the total work needed to form diamond, the sum of the compressive work and the work to induce out of plane strain, a local minimum arises at 20$^{\circ}$ out of plane strain. 
As out of plane strain is induced,
it lowers the barrier to begin the formation of diamond, however, at higher levels of strain, defects and local disorder form during phase transformation, resulting in more compressive work (relative to the work to start transformation) being needed to complete
the phase transformation, causing slower kinetics. This directly differs from previous mechanochemical studies in which increased mechanochemcial work decreases the amount of total work needed to induce reaction \cite{Hamilton2022ManyBody,Hamilton2022Extemporaneous}. 
At the 20$^{\circ}$ local minimum, almost no pure twin defects form, and at higher 
levels of out of plane strain, the uncertainty in twin and hexagonal diamond formation increases significantly as a result of the local disorder induced by the out of plane strain. These results stem from the mechanochemical decrease in activation coupling with the local strains' effect
on other material processes slowing the transformation kinetics.
This opens a number of questions into how the various complications of condensed matter chemistry and phase transformations can influence the role of mechanochemistry by altering downstream and related processes in materials.

\begin{acknowledgments}
Funding for this project was provided by the Director’s Postdoctoral Fellowship program, project LDRD 20220705PRD1. Partial funding was provided by the Advanced Simulation and Computing Physics and Engineering Models project (ASC-PEM). This research used resources provided by the Los Alamos National Laboratory Institutional Computing Program. This work was supported by the U.S. Department of Energy (DOE) through the Los Alamos National Laboratory. The Los Alamos National Laboratory is operated by Triad National Security, LLC, for the National Nuclear Security Administration of the U.S. Department of Energy (Contract No. 89233218CNA000001). Approved for unlimited release: LA-UR-23-21258
\end{acknowledgments}

\includegraphics[scale=0.5,page=1]{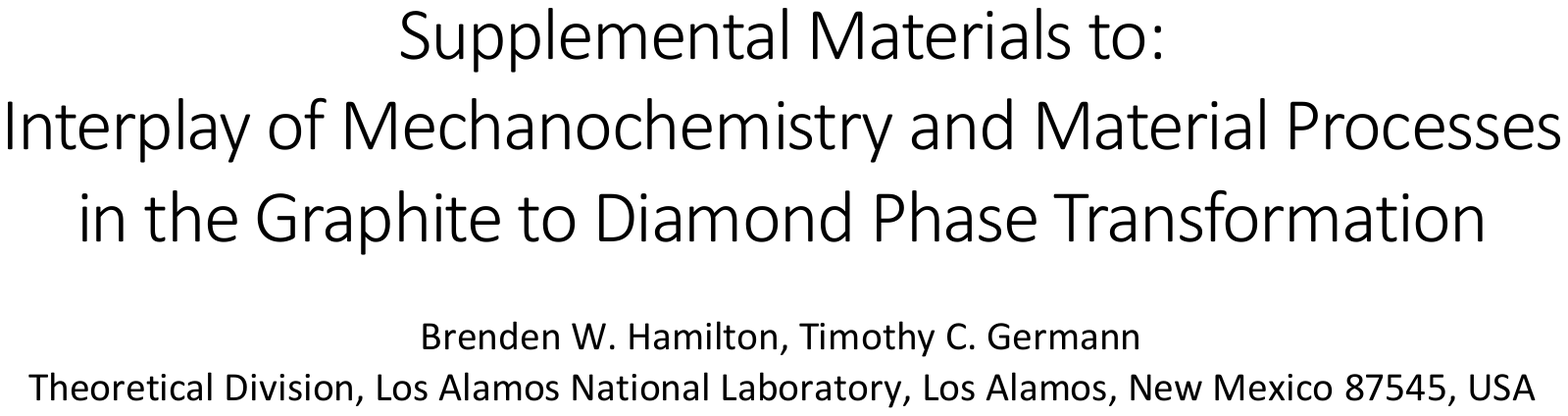}
\includegraphics[scale=0.5,page=2]{Graphite_Supplemental_Materials.pdf}
\includegraphics[scale=0.5,page=3]{Graphite_Supplemental_Materials.pdf}
\includegraphics[scale=0.5,page=4]{Graphite_Supplemental_Materials.pdf}
\bibliography{references}

\end{document}